\documentclass[aps,pre,10pt,twocolumn,superscriptaddress,longbibliography,floatfix]{revtex4-2}

\usepackage[T1]{fontenc}
\usepackage{amsmath,amssymb,bm}
\usepackage{graphicx}
\usepackage{booktabs}

\newcommand{\ket}[1]{\lvert #1\rangle}

\newcommand{\braket}[2]{\langle #1\vert #2\rangle}
\newcommand{\norm}[1]{\left\lVert #1\right\rVert}

\newcommand{\ii}{\mathrm{i}}
\newcommand{\e}{\mathrm{e}}

\newcommand{\CC}{\mathbb{C}}

\begin{document}

\title{Unveiling Semiclassical Structures in Quantum Chaotic Eigenstates Using Neural Networks}

\author{J. Montes}
\email{jmontes.3@alumni.unav.es}
\affiliation{Grupo de Sistemas Complejos, ETSIME, Universidad Polit\'ecnica de Madrid,
Rios Rosas 21, 28003 Madrid, Spain}

\author{F. Borondo}
\email{f.borondo@uam.es}
\affiliation{Departamento de Qu\'imica, Universidad Aut\'onoma de Madrid,
Cantoblanco, 28049 Madrid, Spain}

\author{Gabriel G. Carlo}
\email{g.carlo@conicet.gov.ar}
\affiliation{Comisi\'on Nacional de Energ\'ia At\'omica, CONICET,
Departamento de F\'isica, Av.\ del Libertador 8250,
1429 Buenos Aires, Argentina}

\begin{abstract}
Physics-informed neural networks and neural quantum states have consolidated a new
paradigm to analyze and discover physical phenomena through constrained neural
parametrizations. In this context, we investigate
whether the semiclassical structure of the eigenfunctions of a quantum chaotic
system can be unveiled through unsupervised learning. To this end, we train a "quantum
dictionary", formulated as an overcomplete autoencoder, that sparsely represents the
eigenstates of the system, using as an illustration the quantum baker map. The only explicit physical information imposed
on the dictionary atoms is their localization in phase space, without providing any kind of 
information about the periodic orbits of the corresponding classical system. The model achieves high
fidelity in reconstructing eigenstates not used during training. By comparing the
learned atoms with an independently constructed "semiclassical dictionary", we find
that they spontaneously localize on the periodic orbits and develop scar-like
structures. This result is interesting in two ways: a localization constraint is sufficient to
recover nontrivial semiclassical organization from spectral data and at the same time periodic orbits confirm their fundamental role in the structure of quantum chaotic eigenfunctions. More generally, our proposed architecture opens a new route to learning
representations whose atoms optimize other chosen physical properties.
\end{abstract}

\maketitle

\section{Introduction}

Machine learning is increasingly used in the physical sciences not only as a
high-capacity interpolator, but also as a parametrization whose solutions can
provide information about the physical phenomena under study. Physics-informed
neural networks (PINN) introduced a canonical realization of this principle by
incorporating governing differential equations and boundary conditions into the
training objective \cite{Raissi2019,Karniadakis2021,BorondoRC}. The broader scientific machine
learning program also incorporates physical information through architectures,
coordinates, symmetries, conservation laws, and simple, interpretable models.
Sparse identification of nonlinear dynamics, for example, searches for compact
evolution equations from data \cite{Brunton2016}, while autoencoder-based extensions
learn coordinates in which those dynamics become sparse and interpretable
\cite{Champion2019,Brunton2020}. These works establish a useful principle for the
present problem: physical content can be recovered not by increasing network depth,
but by restricting what the trainable representation can express.

A parallel development has taken place in quantum many-body physics. Neural quantum
states (NQS) use neural networks to parametrize wave-function amplitudes and optimize
them variationally. The seminal restricted-Boltzmann-machine construction of
Carleo and Troyer \cite{CarleoTroyer2017} was followed by applications to strongly
correlated spin and fermionic systems \cite{Nomura2017}, exact and systematically
improvable deep representations \cite{CarleoNomuraImada2018}, neural quantum-state
tomography \cite{Torlai2018}, symmetry-adapted excited states \cite{Choo2018}, and
autoregressive architectures with exact sampling \cite{Sharir2020}. Recent reviews
survey the resulting family of variational, dynamical, and data-driven quantum-state
representations \cite{Lange2024,Medvidovic2024}.

The model developed here lies at the intersection of these two programs, but it is
important to state the relationship precisely. It is not a conventional PINN: no
differential-equation residual is minimized. Nor is it a conventional many-body NQS
that maps a configuration of physical degrees of freedom to a wave-function
amplitude. Instead, the input is a complete state vector
$\ket{\psi}\in\CC^D$, and the network learns a common overcomplete dictionary of
quantum states that analyzes and reconstructs a set of such vectors. The trainable
parameters are therefore themselves normalized wave functions. The dictionary of normalized complex quantum states serves for both the analysis
and reconstruction of eigenstates through a sparse representation. Training
combines a global-phase-insensitive reconstruction loss with regularizers that
reduce mutual overlaps between atoms and favor their phase-space localization. The model
thus constitutes a neural parametrization of quantum states guided by physical
constraints.

This design is based on the principles of sparse coding and dictionary learning. An
overcomplete dictionary can represent each observation using only a few active
atoms, while allowing the atoms themselves to adapt to recurrent structure in the
data. For natural images, this approach has been shown to learn elements that
identify local patterns, such as edges with different orientations, without
introducing this information beforehand \cite{OlshausenField1996}. Subsequent work
developed different methods for learning such dictionaries and relating them to
neural models \cite{Aharon2006,Mairal2010,GregorLeCun2010}. Our construction
transfers this logic to complex quantum states and replaces generic locality with
localization in quantum phase space. The central question here is therefore not only
whether the dictionary reconstructs unseen states, but whether its atoms acquire a
recognizable dynamical meaning.

Quantum chaos is a good arena for testing. Although eigenstates of a classically
chaotic system are broadly extended, their phase-space structure is not
featureless. Heller's discovery of quantum scars showed anomalous concentration of 
quantum probability 
around short unstable periodic orbits \cite{Heller1984}. Semiclassical descriptions
were developed through smoothed wave functions, phase-space constructions, and
linear and nonlinear recurrence theory
\cite{Bogomolny1988,Berry1989,Borondo1994,KaplanHeller1998}. The short-periodic-orbit approach
\cite{VerginiCarlo2000,VerginiCarlo2001,CarloVerginiLustemberg2002} then produced scar functions localized around the hyperbolic structures associated
with classical trajectories. These functions
have been used to construct efficient semiclassical basis sets for calculating
chaotic eigenstates and molecular vibrational states
\cite{Revuelta2013,Revuelta2017,Revuelta2020}. 

The natural next step in our context is to ask whether the localized building blocks emerging from barebone training (just localization is imposed) can themselves be recognized as semiclassical objects. After training, we construct a scar-function dictionary independently and compare it directly with the learned atoms in Hilbert space. We find that the learned atoms display phase-space structures
similar to independently constructed scar functions and localize around their
associated periodic orbits. In addition, both dictionaries show similar median
projection participation over the eigenstate ensemble.

The paper is organized as follows. Section~\ref{sec:baker} defines the classical and
quantum baker maps used as illustration in our study. Section~\ref{sec:scars} presents the construction of periodic
orbits and scar states used for comparison with the learned atoms.
Section~\ref{sec:network} presents the neural quantum dictionary, the loss functions,
and the training and validation protocol. Section~\ref{sec:results} presents the
reconstructions and atom--scar comparisons. Section~\ref{sec:discussion} places the
result within physics-informed machine learning, NQS, sparse coding, and
semiclassical theory, and establishes the limits of the present evidence.

\section{Classical and quantum baker maps}
\label{sec:baker}

\subsection{Classical baker map}

The baker map is an ideal testebed for our method. 
Its classical dynamics is uniformly hyperbolic, its periodic orbits possess an exact
binary symbolic coding, and its quantum version retains the interference and
symmetry phenomena of more general chaotic systems
\cite{BalazsVoros1989,Saraceno1990}. Also, extensive calculations can be carried out 
involving moderate computational load. An explicit periodic-orbit basis is known 
for the quantum baker map \cite{ErmannSaraceno2008}. More broadly,
physics-aware neural architectures have already proved useful for identifying and
quantifying localization in quantum phase space \cite{Montes2025}. 

Working on the unit torus, $(q,p)\in[0,1)\times[0,1)$, the baker map is
\begin{equation}
 B(q,p)=
 \left(
  2q-\nu,\,
  \frac{p+\nu}{2}
 \right),
 \qquad
 \nu=\lfloor 2q\rfloor\in\{0,1\}.
\label{eq:classical-baker}
\end{equation}
That is, it stretches by a factor two along $q$, contracts by a factor two along $p$, and
stacks the right half of the square above the left half. Its Lyapunov exponent per
iteration is
\begin{equation}
 \lambda_{\mathrm{L}}=\ln 2.
\end{equation}

At each iteration, the symbol $\nu_j=\lfloor 2q_j\rfloor$ indicates which of the
two branches of the map is visited by the trajectory. A finite binary string
$\bm{\nu}=\nu_0\nu_1\cdots\nu_{L-1}$ determines a region of phase space if its
continuation is not specified. When repeated periodically, however, it determines
a period-$L$ point. In the convention used in the numerical construction, this
point is
\begin{equation}
 q_0=\frac{(\bm{\nu})_2}{2^L-1},
 \qquad
 p_0=\frac{(\bm{\nu}^{\mathrm{rev}})_2}{2^L-1},
\label{eq:periodic-point}
\end{equation}
where $(\cdot)_2$ denotes the integer represented by the binary string and
$\bm{\nu}^{\mathrm{rev}}$ is the reversed string. The remaining points
$x_j=(q_j,p_j)$ are obtained by applying Eq.~\eqref{eq:classical-baker}.
Cyclic permutations label the same orbit. The scar generator retains only
primitive strings and represents each cyclic class by its lexicographically
smallest rotation.

\subsection{Antiperiodic quantization}

Let $D$ be an even integer and let $\{\ket{q_j}\}_{j=0}^{D-1}$ be the discrete position basis,
with antiperiodic grid points
\begin{equation}
 q_j=\frac{j+\tfrac12}{D}.
\end{equation}
We set the torus area and Planck constant so that
$2\pi\hbar=1/D$. The antiperiodic discrete Fourier matrix is
\begin{equation}
 (G_D)_{jk}
 =
 \frac{1}{\sqrt D}
 \exp\left[
 -\frac{2\pi\ii}{D}
 \left(j+\frac12\right)
 \left(k+\frac12\right)
 \right].
\label{eq:fourier}
\end{equation}
The Balazs--Voros--Saraceno quantum baker operator is
\begin{equation}
 U_B
 =
 G_D^\dagger
 \begin{pmatrix}
  G_{D/2} & 0\\
  0 & G_{D/2}
 \end{pmatrix}.
\label{eq:quantum-baker}
\end{equation}
It is unitary and quantizes the two classical branches while preserving the
parity symmetry afforded by antiperiodic boundary conditions
\cite{BalazsVoros1989,Saraceno1990}.

The eigenstates and eigenphases are defined by
\begin{equation}
 U_B\ket{\psi_n}=\e^{-\ii\varphi_n}\ket{\psi_n},
 \qquad
 n=0,\ldots,D-1,
\label{eq:eigenproblem}
\end{equation}
Since $U_B$ is unitary, its eigenstates can be chosen normalized and mutually
orthogonal. Each eigenstate is defined up to an arbitrary global phase, which does
not affect physical observables.

\subsection{Phase-space representation}

To represent a quantum state in phase space, we calculate its Husimi density from
projections onto coherent states centered at each point $(q,p)$. Since phase space
is a torus with antiperiodic boundary conditions, the Gaussian packet must be
make periodic. In the position basis, the toroidal coherent state takes the form
\begin{align}
 \braket{q_j}{q,p}_{\mathrm{tor}}
 ={}&
 \mathcal{N}_{q,p}
 \sum_{m\in\mathbb{Z}}
 \exp\left[-\pi D(q_j+m-q)^2\right]
 \nonumber\\
 &{}\times
 \exp\left[
 2\pi\ii D\left(q_j+m-\frac{q}{2}\right)p-\ii\pi m
\right],
\label{eq:torus-coherent}
\end{align}
where $\mathcal{N}_{q,p}$ is the normalization constant. The sum incorporates the
periodic images of a single packet centered at $(q,p)$. The Husimi density of a
normalized state $\ket{\phi}$ is then defined as
\begin{equation}
 H_\phi(q,p)=\left|\braket{q,p_{\mathrm{tor}}}{\phi}\right|^2.
\label{eq:husimi}
\end{equation}
We use this definition throughout all our study.

\section{Semiclassical scar dictionary}
\label{sec:scars}

The scar dictionary is generated independently of the neural optimization. Its
construction follows the short-periodic-orbit philosophy of
Refs.~\cite{VerginiCarlo2000,VerginiCarlo2001,CarloVerginiLustemberg2002,
ErmannSaraceno2008,Revuelta2013,Revuelta2020}: first build an
orbit-centered tube state, then perform a forward and backward
propagation over a finite time.

\subsection{Periodic-orbit modes}

For the point $x_j=(q_j,p_j)$ of a period-$L$ orbit, define the one-step generating
action used by the implementation,
\begin{equation}
 S_j=\nu_j\left(\frac{p_j}{4}+\frac{q_j}{2}+\frac14\right),
 \qquad
 \nu_j=\lfloor 2q_j\rfloor,
\label{eq:local-action}
\end{equation}
and the total orbit action
\begin{equation}
 S_{\bm{\nu}}=\sum_{j=0}^{L-1}S_j.
\end{equation}
For $k=0,\ldots,L-1$, the quasienergy associated with a periodic-orbit mode is
\begin{equation}
 \epsilon_{\bm{\nu},k}
 =
 \frac{2\pi}{L}\left(DS_{\bm{\nu}}+k\right).
\label{eq:scar-quasienergy}
\end{equation}
Writing
\begin{equation}
 \theta_j=2\pi D\sum_{\ell=0}^{j-1}S_\ell,
 \qquad \theta_0=0,
\end{equation}
the unnormalized periodic-orbit mode (POM), or tube function, is
\begin{align}
 \ket{\phi_{\bm{\nu}}^k}
 ={}&
 \frac{1}{\sqrt L}
 \sum_{j=0}^{L-1}
 \exp\left[
 -\frac{2\pi\ii}{L}(DS_{\bm{\nu}}+k)j
 \right]
 \nonumber\\
 &{}\times\e^{-\ii\theta_j}\ket{q_j,p_j}_{\mathrm{tor}}.
\label{eq:pom}
\end{align}
The state is normalized numerically after the coherent packets are summed.

\subsection{Discrete symmetries}

The baker map has reflection and time-reversal symmetries. In the numerical
representation used here they act as
\begin{equation}
 R=-G_DG_D,
 \qquad
 T=G_D\mathcal{K},
\label{eq:symmetries}
\end{equation}
where $\mathcal{K}$ denotes complex conjugation in the position basis. Reflection
maps a binary string to its bitwise complement, while time reversal reverses the
order of the string. Let $\sim$ denote equality up to a cyclic rotation and abbreviate
$\ket{\phi}=\ket{\phi_{\bm{\nu}}^k}$. The code groups the four symbolic classes
$\bm{\nu}$, $R\bm{\nu}$, $T\bm{\nu}$, and $RT\bm{\nu}$, keeps the first
lexicographic representative encountered, and then constructs
\begin{equation}
\begin{array}{c|c}
\text{cyclic-class relations}&\text{POM candidates}\\
\hline
R\bm{\nu}\sim\bm{\nu},\ T\bm{\nu}\sim\bm{\nu}
 & \ket{\phi}\\
R\bm{\nu}\sim\bm{\nu},\ T\bm{\nu}\not\sim\bm{\nu}
 & \ket{\phi}\pm T\ket{\phi}\\
R\bm{\nu}\not\sim\bm{\nu},\ T\bm{\nu}\sim\bm{\nu}
 & \ket{\phi}\pm R\ket{\phi}\\
R\bm{\nu}\not\sim\bm{\nu},\ T\bm{\nu}\not\sim\bm{\nu}
 & (1\pm R)(1\pm T)\ket{\phi}.
\end{array}
\label{eq:symmetry-cases}
\end{equation}
Each candidate is normalized separately. Equation~\eqref{eq:symmetry-cases} describes
the implemented branching literally; in particular, cyclic permutations are discarded.

\subsection{Finite-time scar functions}

Each POM candidate produced by Eq.~\eqref{eq:symmetry-cases} is propagated and this is modulated by a Gaussian time window:
\begin{align}
 \ket{f_{\bm{\nu},k}}
 \propto {}&
 \ket{\phi_{\bm{\nu}}^k}
 \nonumber\\
 &+
 \sum_{\ell=1}^{\ell_{\max}}
 \e^{-\ell^2/(2t_{\mathrm{E}}^2)}
 \e^{-\ii\epsilon_{\bm{\nu},k}\ell}
 U_B^\ell\ket{\phi_{\bm{\nu}}^k}
 \nonumber\\
 &+
 \sum_{\ell=1}^{\ell_{\max}}
 \e^{-\ell^2/(2t_{\mathrm{E}}^2)}
 \e^{+\ii\epsilon_{\bm{\nu},k}\ell}
 U_B^{-\ell}\ket{\phi_{\bm{\nu}}^k}.
\label{eq:scar-filter}
\end{align}
We use
\begin{equation}
 t_{\mathrm{E}}=\frac{\ln D}{2\ln 2},
 \qquad
 \ell_{\max}=20,
\label{eq:scar-times}
\end{equation}
and normalize Eq.~\eqref{eq:scar-filter}. The time scale is one half of the
logarithmic Ehrenfest time in map iterations. Primitive orbit classes are visited
in increasing period. Within each period, the strings are ordered by their binary
value and, for each one, increasing values of $k=0,\ldots,L-1$ are considered.
Generation stops immediately when
\begin{equation}
 M_{\mathrm{scar}}=\operatorname{round}(1.2D)=290
\end{equation}
states have been appended. All scar functions generated by orbits up to period seven are used, whereas
period eight is only partially considered.

\section{Physics-informed neural quantum dictionary}
\label{sec:network}

\begin{figure*}[t]
 \includegraphics[width=\textwidth]{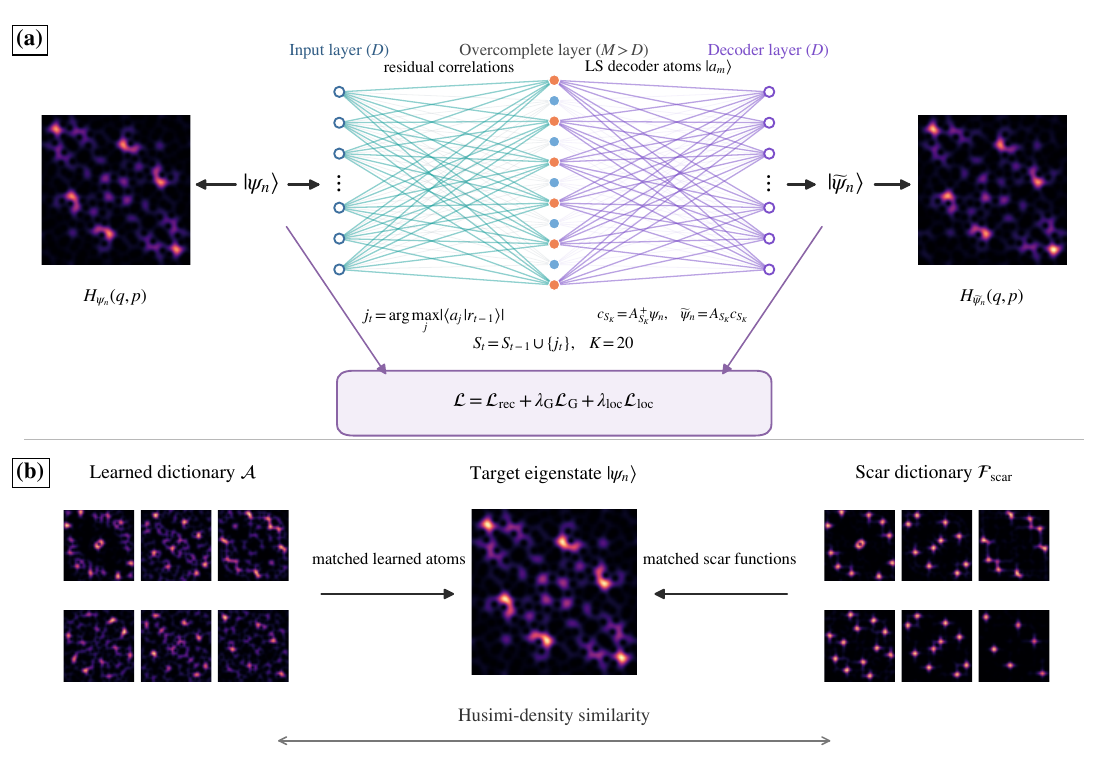}
 \caption{\label{fig:workflow}
 Architecture and dictionary-comparison scheme. (a) A normalized
 quantum-baker eigenstate is represented by its complex components and its Husimi
 density. The sparse dictionary network
 has widths $D\to M\to D$, with $M>D$.
 The active support is selected by residual correlations, as in complex
 orthogonal matching pursuit, and the decoder reconstructs the state by a
 least-squares recombination of the selected atoms. Training combines phase-insensitive reconstruction,
 dictionary-overlap diversity, and phase-space-localization losses. (b) The learned and
 independently constructed scar dictionaries provide two overcomplete
 representations of the same eigenstate.
 The six elements displayed on each side are visually matched atom--scar
 pairs selected after training by Husimi-density morphology. After training, both dictionaries are
 compared through $O_{mr}=|\langle a_m|f_r^{\rm scar}\rangle|^2$. }
\end{figure*}

Our objective is to learn a set of functions capable of representing the
eigenstates while remaining individually interpretable. We therefore use a neural
architecture in the form of a dictionary: each unit in the latent space is
associated with a quantum state, which we call an atom. 
For each input state, the encoder first projects the state onto the
learned atoms, giving the initial overlaps $z_m=\langle a_m|\psi\rangle$.
These overlaps are used to start a sparse residual-selection procedure; after
each selected atom, the decoder recomputes all active coefficients by complex
least squares. The trainable wave functions are still a single learned dictionary
shared by all eigenstates.

The architecture is an overcomplete autoencoder because the latent dimension $M$
is larger than the input-state dimension $D$. Consequently, the atoms do not form
a basis in the strict sense, but an overcomplete dictionary that represents each
state using only a few active elements. The term \emph{neural} indicates that this
dictionary is learned by training a neural network, while its quantum and
physics-informed character comes from working with complex states and incorporating
reconstruction, phase-space localization, and control of the mutual overlaps between atoms into the loss function. This construction is related to sparse coding and
dictionary learning \cite{Aharon2006,Mairal2010,GregorLeCun2010}, within the broader
framework of neural representations of quantum states
\cite{CarleoTroyer2017,Lange2024,Medvidovic2024}.

\subsection{Overcomplete dictionary and OMP--LS sparse inference}
Here OMP denotes orthogonal matching pursuit and LS denotes least squares.
OMP is a greedy sparse-approximation method: it selects one dictionary element at
a time using its correlation with the current residual, and then orthogonalizes
the approximation by solving the least-squares problem on the active support
\cite{Pati1993,TroppGilbert2007}.

The input is a normalized state $\ket{\psi}\in\CC^D$. Let
$a_{mj}=\braket{q_j}{a_m}$ be the component of atom $\ket{a_m}$ in the position
basis. The dictionary is represented by the matrix
\begin{equation}
 A=
 \begin{pmatrix}
  \bm a_1^{\mathsf T}\\
  \vdots\\
  \bm a_M^{\mathsf T}
 \end{pmatrix}
 \in\CC^{M\times D},
 \qquad
 \bm a_m=
 \frac{\bm b_m}{\norm{b_m}_2}.
\label{eq:dictionary}
\end{equation}
where the complex vectors $\bm b_m$ contain the trainable parameters and each atom
$\bm a_m$ is kept normalized. In the model considered here,
$M=290\simeq1.20D$, so the
dictionary is overcomplete.

The first encoder operation is the same projection used in the earlier
architecture: the input eigenstate is projected onto all atoms. In the present
architecture these initial overlaps are not kept as the final coefficients; they
only determine the first active atom. After that first reconstruction, the same
projection operation is applied to the residual. For a generic vector $r$, the
complex correlation is
\begin{equation}
 z_m(r)=\braket{a_m}{r}
 =\sum_{j=0}^{D-1}a_{mj}^*r_j.
\label{eq:analysis}
\end{equation}

In the first step $r_0=\psi$, and in later steps the same formula is applied
with $\psi$ replaced by the residual $r_{t-1}$.

\subsection{Residual OMP support and least-squares decoder}

Let $S_0=\varnothing$ and $r_0=\psi$. At iteration $t=1,\ldots,K$ the
support is enlarged by
\begin{equation}
 j_t=\arg\max_{j\notin S_{t-1}}
 \left|\braket{a_j}{r_{t-1}}\right|,
 \qquad
 S_t=S_{t-1}\cup\{j_t\}.
\label{eq:omp-support}
\end{equation}

After every support update, all active coefficients are recomputed by the
complex least-squares problem
\begin{equation}
 \bm c_{S_t}
 =
 \arg\min_{\bm c}
 \left\|
  \ket{\psi}
  -\sum_{j\in S_t}c_j\ket{a_j}
 \right\|_2^2,
 \qquad
 r_t=\ket{\psi}-\sum_{j\in S_t}c_j\ket{a_j}.
\label{eq:omp-ls}
\end{equation}

Equivalently, if $A_{S_t}$ contains the selected atoms as columns, then
$\bm c_{S_t}=A_{S_t}^{+}\psi$ up to the small numerical ridge used in the solve. The
final decoder is
\begin{equation}
 \ket{\widetilde{\psi}}
 =
 \sum_{j\in S_K}c_j\ket{a_j}.
\label{eq:decoder}
\end{equation}
Equation~\eqref{eq:decoder} is not an orthogonal projector because the dictionary is
overcomplete and generally nonorthogonal. The latent representation is therefore
restricted to 
$K$ complex coefficients chosen by residual matching and jointly debiased
by least squares, imposing a sparse representation of each state.
Figure~\ref{fig:workflow} summarizes the complete computation.

\subsection{Phase-insensitive reconstruction loss}

Before comparing the original state and its reconstruction, both are normalized:
\begin{equation}
 \ket{\bar\psi}=\frac{\ket{\psi}}{\norm{\psi}_2},
 \qquad
 \ket{\bar{\widetilde{\psi}}}
 =\frac{\ket{\widetilde{\psi}}}{\norm{\widetilde{\psi}}_2}.
\end{equation}
Since two states that differ only by a global phase represent the same physical
state, we align the phase of the reconstruction through
\begin{equation}
 \gamma=\arg\braket{\bar\psi}{\bar{\widetilde{\psi}}},
 \qquad
 \ket{\psi_{\mathrm{al}}}
 =\e^{-\ii\gamma}\ket{\bar{\widetilde{\psi}}}.
\label{eq:phase-align}
\end{equation}
The reconstruction loss is the mean squared distance between both states after
phase alignment:
\begin{equation}
 \mathcal{L}_{\mathrm{rec}}
 =
 \frac{1}{|\mathcal{B}|D}
 \sum_{\psi\in\mathcal{B}}
 \norm{\ket{\psi_{\mathrm{al}}}-\ket{\bar\psi}}_2^2.
\label{eq:rec-loss}
\end{equation}
The fidelity between the original state and its reconstruction is
\begin{equation}
 F(\psi,\widetilde{\psi})
 =
 \frac{|\braket{\psi}{\widetilde{\psi}}|^2}
 {\braket{\psi}{\psi}\braket{\widetilde{\psi}}{\widetilde{\psi}}}.
\label{eq:fidelity}
\end{equation}
For each state, the loss and fidelity are related by
\begin{equation}
\mathcal{L}_{\mathrm{rec}}(\psi)
=\frac{2}{D}\left(1-\sqrt{F(\psi,\widetilde{\psi})}\right).
\label{eq:loss-fidelity-relation}
\end{equation}
Thus, minimizing the loss is equivalent to maximizing the fidelity, independently
of the global phase chosen to represent the states.
\subsection{Gram-overlap regularization}

Let
\begin{equation}
 G_{mn}=\braket{a_m}{a_n}
 \qquad
 (m,n=1,\ldots,M)
\end{equation}
be the Gram matrix of the dictionary. The Gram-overlap loss is defined as
\begin{equation}
 \mathcal{L}_{\mathrm{G}}
 =
 \frac{1}{M^2}\norm{G-I_M}_{F}^{2}
 =
 \frac{1}{M^2}
 \sum_{m,n=1}^{M}
 \left|G_{mn}-\delta_{mn}\right|^2.
\label{eq:gram-loss}
\end{equation}
It discourages duplicate or nearly parallel atoms. It does not impose an
orthonormal basis: when $M>D$, $M$ mutually orthogonal nonzero vectors in $\CC^D$
cannot exist. Indeed, the Welch bound implies
\begin{equation}
 \mathcal{L}_{\mathrm{G}}\geq \frac{1}{D}-\frac{1}{M}
\label{eq:welch}
\end{equation}
for unit-norm atoms \cite{Welch1974}. The term should therefore be interpreted as a
Gram-overlap diversity regularizer for an overcomplete frame.

\subsection{Phase-space localization regularization}

To favor localized atoms, we evaluate their Husimi density over a set of coherent
states $\{\ket{q_\alpha,p_\alpha}\}$. For atom $a_m$, we define
\begin{equation}
 h_{m\alpha}
 =
 \left|\braket{q_\alpha,p_\alpha}{a_m}\right|^2.
\end{equation}
The participation number of this distribution is
\begin{equation}
 P_{\mathrm{H}}(a_m)
 =
 \frac{\left(\sum_\alpha h_{m\alpha}\right)^2}
 {\sum_\alpha h_{m\alpha}^2}.
\label{eq:husimi-participation}
\end{equation}
Let $P_{\mathrm{coh}}$ be the corresponding value for a coherent state, which sets
the reference localization scale. The normalized functional and its average over
the dictionary are
\begin{equation}
 \mu(a_m)=\frac{P_{\mathrm{H}}(a_m)}{P_{\mathrm{coh}}},
 \qquad
 \mathcal{L}_{\mathrm{loc}}
 =
 \frac1M\sum_{m=1}^{M}\mu(a_m).
\label{eq:loc-loss}
\end{equation}
Minimizing $\mathcal{L}_{\mathrm{loc}}$ favors phase-space support comparable to
that of a coherent state, without introducing information about periodic orbits.

\subsection{Objective function and training protocol}

The total objective function combines the three contributions above:
\begin{equation}
 \mathcal{L}
 =
 \mathcal{L}_{\mathrm{rec}}
 +\lambda_{\mathrm{G}}\mathcal{L}_{\mathrm{G}}
 +\lambda_{\mathrm{loc}}\mathcal{L}_{\mathrm{loc}}.
\label{eq:total-loss}
\end{equation}
The weights $\lambda_{\mathrm{G}}$ and $\lambda_{\mathrm{loc}}$ control,
respectively, the diversity and localization of the atoms relative to
reconstruction quality. The main model parameters are summarized in
Table~\ref{tab:hyperparameters}.

For $D=242$, 218 eigenstates are used for training and 24 for validation. The
latter do not participate in the optimization and allow us to evaluate the ability
of the dictionary to reconstruct states not used during learning.

\begin{table}[!b]
\caption{\label{tab:hyperparameters}
Main parameters of the model and the semiclassical comparison.}
\begin{ruledtabular}
\begin{tabular}{lc}
Quantity & Value\\
\hline
Hilbert-space dimension $D$ & 242\\
Learned atoms $M$ & 290\\
Active atoms $K$ & 20\\
Training / validation states & 218 / 24\\
$\lambda_{\mathrm{G}}$ & $10^{-2}$\\
$\lambda_{\mathrm{loc}}$ & $3.0\times10^{-8}$\\
Sparse inference rule & OMP residual selection + LS\\
Scar functions $M_{\mathrm{scar}}$ & 290\\
Largest orbit period reached & 8\\
Scar propagation cutoff $\ell_{\max}$ & 20
\end{tabular}
\end{ruledtabular}
\end{table}

The validation set was also used to select some model parameters. It therefore
measures behavior outside the training set, but does not constitute a completely
independent test set.

\section{Results}
\label{sec:results}

The main result of this work is that the proposed architecture learns a
representation of the eigenstates in terms of a common set of atoms whose
properties can be controlled during training. In this case, in addition to
requiring the atoms to reconstruct the eigenstates using only a few active
elements, we favor their localization in phase space. Scar functions are also
localized states, but are constructed from periodic orbits. This connection
motivates the analysis below: we first verify that the learned dictionary
adequately reconstructs the eigenstates and then investigate whether the localized
atoms obtained by the network display a structure similar to scar functions.

\subsection{Sparse reconstruction of unseen eigenstates}

\begin{figure}[!t]
 \includegraphics[width=\columnwidth]{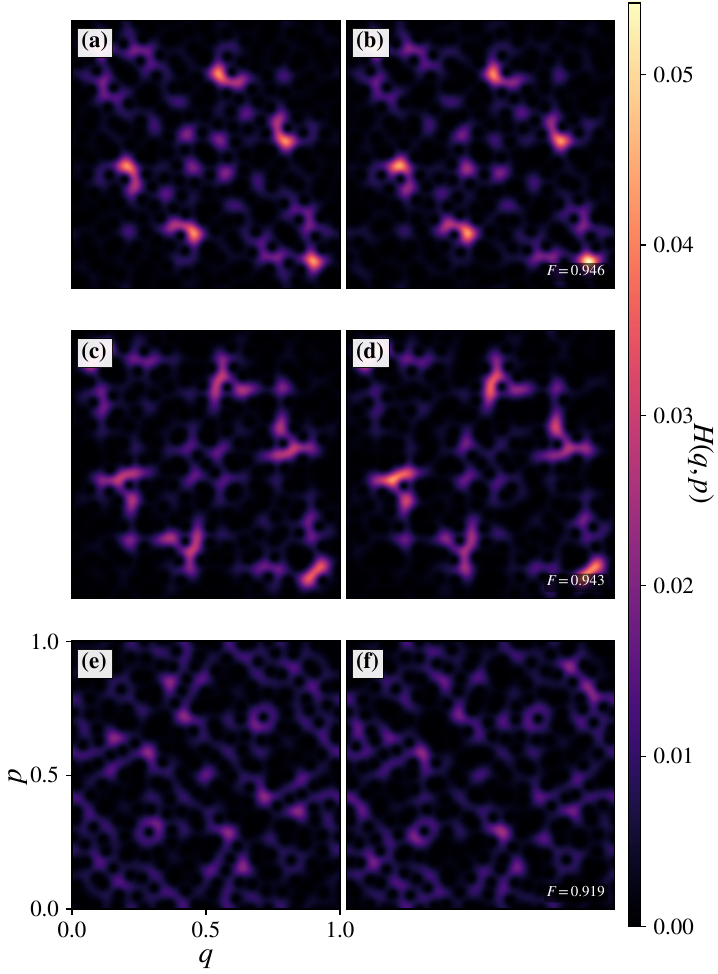}
 \caption{\label{fig:reconstruction}
 Husimi densities of three validation eigenstates (left column) and their
 OMP+least-squares sparse reconstructions
 (right column). All six panels use one common color scale.
 The selected state indices are 63, 208, and 232;
 their fidelities are
 $0.9456$, $0.9431$, and $0.9187$,
 respectively. Only $K=20$ of the $M=290$
 learned atoms contribute to each reconstruction.}
\end{figure}

Once trained, the network reconstructs with high fidelity eigenstates that were not
used during learning. The mean fidelity for these states is
\begin{equation}
 \overline F_{\mathrm{val}}=0.8764100,
\end{equation}
with sample standard deviation $0.0584644$, median
$0.8914293$, minimum
$0.6786816$, and maximum
$0.9455745$.  Most validation states are reconstructed
with high fidelity, while the larger dispersion identifies a small number of
harder states for this fixed value of $K$. The learned atom dictionary therefore does
not merely represent the training states, but generalizes to new eigenstates.

Figure~\ref{fig:reconstruction} illustrates this ability for eigenstates with very
different structures. 
The first two validation examples, shown in panels (a)--(b) and
(c)--(d), are localized in phase space, whereas the third example, panels
(e)--(f), displays a more extended Husimi distribution.
Although the dictionary atoms are constrained to be localized, their sparse
combinations correctly reconstruct both kinds of states, as shown in panels (b),
(d), and (f). This result suggests that the global extension of a chaotic
eigenstate can arise from the superposition of a small number of localized
structures. As we show in the next subsection, these structures organize around
short periodic orbits, indicating that such orbits are relevant components of the
eigenfunctions even when the latter appear delocalized in phase space.

\subsection{Comparison between learned atoms and scar functions}

\begin{figure}[!t]
 \includegraphics[width=\columnwidth]{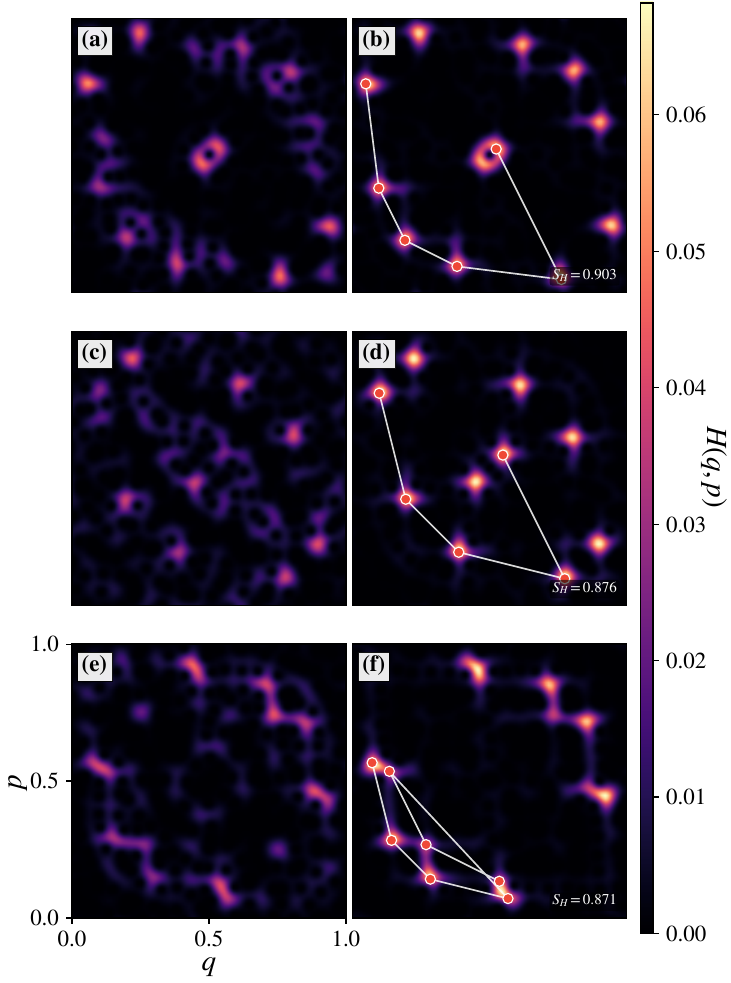}
 \caption{\label{fig:atom-scar}
 Three learned atoms (left column) and the independently generated scar function
 selected after training by Husimi-density
 morphology, with substantial Hilbert-space overlap (right column). The associated periodic orbit is
 overlaid on each scar Husimi density. From top to bottom, the pairs are
 $(m,r)=(90,69)$, $(128,37)$, and $(147,186)$, with squared overlaps
 $0.6659$, $0.5057$, and $0.6893$. The binary strings labeling the orbits are
 $000011$, $00011$, and $0001001$, respectively.  Their Husimi-density similarities are $0.903$, $0.876$,
 $000011$, $00011$, and $0001001$, respectively. Their Husimi-density similarities are $0.903$, $0.876$,
 and $0.871$, respectively.}
\end{figure}

For every learned atom and scar function we compute
\begin{align}
 O_{mr}
 &=
 \left|\braket{a_m}{f_r}\right|^2,
\nonumber\\
 &m=1,\ldots,M,\qquad
 r=1,\ldots,M_{\mathrm{scar}}.
\label{eq:atom-scar-overlap}
\end{align}
No phase-space image processing enters the computation of
the Hilbert-space overlap matrix. The displayed examples in
Fig.~\ref{fig:atom-scar} are selected, after training, by Husimi-density
morphology so that the visual scar-like structures are clear.
Figure~\ref{fig:atom-scar}
shows three high-similarity examples. The recurrent ridges and maxima of each learned
atom align with those of a scar function and with its underlying periodic orbit.
This is nontrivial because periodic-orbit locations, symbolic codes, actions, and
symmetry labels are absent from the neural objective.

The overlap values must be interpreted in light of the different conditions under
which the two dictionaries are constructed. Scar functions are explicitly
symmetrized, and the semiclassical dictionary is built from short periodic orbits.
Consequently, their Husimi densities are strongly concentrated around orbital
structures and have very little intensity away from them, as seen in panels (b)
and (f) and in the central region of panel (d). The learned atoms, by contrast, are
not subject to any symmetry and receive no information about the orbits. They are
only encouraged to be localized and, collectively, to reconstruct the eigenstates.
Thus, although they reproduce the main scar ridges, they may retain fluctuations
outside the localized regions,  In the examples shown here, atoms (c) and
(e) display residual central fluctuations that are weaker or absent in the
corresponding scar functions, while atom (a) has small additional intensity around
the same classical periodic-orbit structure highlighted in panel (b). These differences reduce the global overlap even when
the localized structure of both states is very similar.

Moreover, the correspondence need not be one-to-one. Some atoms may represent only
part of the structure of a scar function, while a combination of several atoms
recovers a similar global shape. This is natural in an overcomplete sparse
dictionary, where information associated with the same classical structure may be
distributed among several elements. From this perspective, squared overlaps close
to $0.5$ are substantial: they emerge without imposing periodic orbits or
symmetries, and despite the differences outside the localized regions.

The full overlap matrix is broad rather than one-to-one. Its largest entry is
0.741091. For each atom, the mean and median of $\max_r O_{mr}$ are
$0.229053$ and $0.200598$,
respectively; 234 of 290 atoms exceed $0.1$,
145 exceed $0.2$, and
70 exceed $0.3$. Conversely, 285 of the 290 scar functions have at least one atom
with overlap above $0.1$. These statistics support the emergence of a distributed
scar-like organization without implying an exact identification between the
elements of the two dictionaries.

\subsection{Global representational efficiency}

\begin{figure*}[t]
 \includegraphics[width=\textwidth]{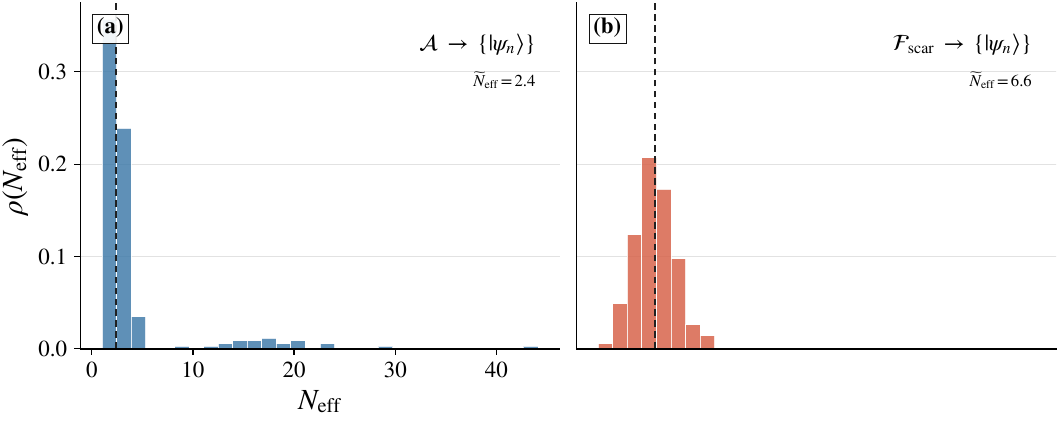}
 \caption{\label{fig:participation}
 Distribution, over all $D=242$ quantum-baker eigenstates, of the effective
 projection participation $N_{\mathrm{eff}}$ for (a) the learned dictionary and
 (b) the independently constructed scar dictionary. Dashed lines mark medians,
 $2.427$ and $6.585$, respectively. The
 corresponding means are $3.991$ and
 $6.603$. Keep in mind that dictionaries are nonorthogonal thus this measure does not reflect an exact expansion.}
\end{figure*}

To compare both dictionaries on the same task without retraining coefficients, let
$\{\ket{d_m}\}$ denote either the set of normalized learned atoms or the set of
normalized scar functions. For each of the $D=242$ eigenstates $\ket{\psi_n}$, we
calculate its projection onto all the elements of each dictionary:
\begin{equation}
 w_{nm}=\left|\braket{d_m}{\psi_n}\right|^2
\end{equation}
and define the effective participation from these weights as
\begin{equation}
 N_{\mathrm{eff}}^{(n)}
 =
 \frac{\left(\sum_m w_{nm}\right)^2}
 {\sum_m w_{nm}^2}.
\label{eq:effective-participation}
\end{equation}
This quantity estimates how many dictionary elements contribute significantly to
the representation of an eigenstate. If a single projection dominates,
$N_{\mathrm{eff}}=1$; if the weight is equally distributed among $r$ elements,
$N_{\mathrm{eff}}=r$. Small values therefore indicate a representation
concentrated on a few elements, whereas large values correspond to a more
distributed representation.

This procedure produces 242 values of $N_{\mathrm{eff}}$ for the learned
dictionary and another 242 for the scar dictionary. Figure~\ref{fig:participation}
represents both collections using normalized histograms. For a direct comparison,
the two panels use the same intervals and vertical scale. The interval width is
chosen by applying the Freedman--Diaconis rule to the union of both collections,
\begin{equation}
 h=2\,\frac{\operatorname{IQR}(N_{\mathrm{eff}})}{N^{1/3}},
\end{equation}
where $\operatorname{IQR}$ is the interquartile range and $N$ is the total number
of values considered \cite{FreedmanDiaconis1981}. The dashed lines indicate the
median of each distribution.

Figure~\ref{fig:participation} shows  that the learned dictionary has a smaller median
projection participation than the scar dictionary for the present optimized
architecture. In particular,  the learned atoms form a more
concentrated projection representation of a large fraction of the eigenstate
ensemble, while the scar dictionary provides a stable semiclassical reference.

The learned distribution still contains a small tail toward larger values of
$N_{\mathrm{eff}}$. One possible explanation is that the atoms are neither
symmetrized nor constructed from a truncated set of short periodic orbits. By
learning directly from the eigenstates, they may indirectly incorporate
contributions from other dynamical structures and from orbits not included in the
scar dictionary. This information may be spread among a larger number of atoms,
producing a somewhat less concentrated representation for some eigenstates. The
tail reflects this difference, while  the main peak at small $N_{\mathrm{eff}}$
shows that the new architecture learns atoms that are highly efficient under this
projection diagnostic.

There are two necessary qualifications. First, the dictionaries have
the same cardinality in the
present run, $290$ learned atoms and $290$ scar functions. Second, for a nonorthogonal overcomplete dictionary,
the projections $\braket{d_m}{\psi_n}$ are analysis coefficients, not unique
expansion coefficients. Equation~\eqref{eq:effective-participation} is therefore a
global concentration diagnostic, not a solution of the minimum-support
representation problem.

\begin{table}[!t]
\caption{\label{tab:results}
Summary statistics for the final learned dictionary. Atom--scar entries are squared Hilbert-space overlaps.}
\begin{ruledtabular}
\begin{tabular}{p{0.64\columnwidth}c}
Quantity & Value\\
\hline
Mean training fidelity & 0.999291\\
Mean validation fidelity & 0.876410\\
Median validation fidelity & 0.891429\\
Validation range & $[0.678682,0.945575]$\\
Largest atom--scar overlap & 0.741091\\
Mean best scar overlap per atom & 0.229053\\
Median best scar overlap per atom & 0.200598\\
Atoms with best overlap $\geq0.1$ & 234 / 290\\
Atoms with best overlap $\geq0.3$ & 70 / 290\\
Median $N_{\mathrm{eff}}$, learned & 2.427\\
Median $N_{\mathrm{eff}}$, scars & 6.585
\end{tabular}
\end{ruledtabular}
\end{table}

\section{Discussion and conclusions}
\label{sec:discussion}
The main result of this work is a way to search for common representations of a
family of eigenstates whose elements, besides reconstructing them, optimize
physical properties chosen in advance. The neural architecture does not act here
as a black box that produces an internal representation that is difficult to
interpret. Its trainable parameters are directly the dictionary atoms, that is,
wave functions that can be visualized in phase space and compared with known
physical constructions.

This interpretability follows from the architecture itself. The sparse support is built iteratively from overlaps with the
current residual, and once the support is selected the decoder recomputes all active
coefficients by least squares. The model therefore learns an overcomplete set shared by all eigenstates,
rather than an independent representation for each one. The high fidelity obtained
for states that did not participate in training shows that the dictionary captures
common structures and does not merely memorize the set used during learning.

In this case, we chose phase-space localization as the property to be favored by
the atoms. The most relevant physical result is that this single condition,
combined with the need to reconstruct the eigenstates sparsely, causes structures
organized around short periodic orbits to emerge. The network is not given the
orbit positions, actions, symbolic codes, or scar-function symmetries. The
similarity therefore appears as a consequence of the structure of the eigenstates
themselves. This reinforces the interpretation of periodic orbits as organizing
elements of chaotic quantum eigenfunctions, even when the latter display extended
phase-space distributions.

The combination of the different constraints is essential. Sparse reconstruction
requires the same atoms to participate recurrently in many eigenstates;
localization prevents those atoms from becoming extended combinations without a
clear interpretation; and Gram-overlap regularization encourages the dictionary to
explore different structures instead of accumulating redundant elements. Together,
these conditions select localized functions that must be collectively useful for
describing a complete family of eigenstates. The detailed morphology of the atoms
is not prescribed, but emerges during learning.

Comparison with scar functions makes it possible to recognize the physical meaning
of this morphology. An exact one-to-one correspondence should not be expected:
scars are symmetrized and constructed from a truncated set of periodic orbits,
whereas the learned atoms are not subject to these conditions. An atom may
represent a deformed scar function, part of one, or a structure combining
contributions associated with different orbits. Despite these differences, the
overlaps, Husimi shapes, and participation distributions show that both dictionaries organize the same semiclassical
structures, while the optimized learned dictionary gives a more concentrated
projection-participation distribution in the present run.

The tail observed in the participation of the learned dictionary also points to a
physically suggestive difference. Because the atoms are not restricted by
symmetrization or a maximum orbital period, they may indirectly incorporate
contributions from a broader set of orbits and dynamical structures. For some
eigenstates, this information can generate a long tail even though the median
participation is smaller than that of the scar dictionary.

This approach has been made possible by combining ideas that usually appear in
different contexts. From physics-informed learning, it takes the principle of
incorporating physical knowledge into the model constraints
\cite{Raissi2019,Karniadakis2021}; from neural quantum states, the parametrization
of wave functions through trainable architectures
\cite{CarleoTroyer2017,Lange2024,Medvidovic2024}; and from dictionary learning,
overcompleteness, sparsity, and the interpretability of each element
\cite{OlshausenField1996,Aharon2006,Mairal2010,GregorLeCun2010}. Short-periodic-orbit
theory finally provides the physical framework with which to identify the
discovered structures \cite{VerginiCarlo2000,Revuelta2013,Revuelta2020}. The
contribution is not simply to bring these tools together, but to use them to turn a
desired physical property into a condition on a directly interpretable neural
representation.

During development, different Hilbert-space dimensions and different
hyperparameter values were explored. These explorations showed that the appearance
of localized atoms and their scar-like organization are not tied to an isolated
parameter choice. The configuration presented here makes it possible to display
the mechanism clearly and to perform a detailed comparison between the two
dictionaries.

The same idea can be extended beyond localization. The objective function could
favor symmetries, prescribed values of observables, concentration on classical
regions, or other properties adapted to the system under study. The architecture
thus provides a general strategy for searching for the building blocks that best
represent a family of states while simultaneously extremizing a quantity with
physical meaning. Applying it to chaotic billiards, molecular vibrational states,
open maps, and many-body phase spaces would allow the learned dictionary to be used
as a tool for discovering and comparing recurrent physical structures.

In summary, we have presented a neural architecture that searches for
interpretable representations of a family of quantum states while controlling the
physical properties of their elements. The network learns a shared overcomplete
dictionary whose atoms are explicit wave functions. By requiring these atoms to
reconstruct the eigenstates sparsely while remaining localized in phase space, we
obtain a compact representation that generalizes to states not used during
training. The central physical result is that imposing localization on the atoms
naturally causes structures organized around short periodic orbits to emerge. The
network receives no information about these orbits or about scar functions.
Nevertheless, the learned atoms reproduce their main Husimi structures, display
substantial overlaps with them, and provide a global representation of the
eigenstates with an effective participation more concentrated than, under the
projection diagnostic, that of the semiclassical dictionary. Even extended
eigenstates can be reconstructed from combinations of these localized elements.
This shows that periodic orbits are a relevant structure in the organization of
chaotic quantum eigenfunctions, beyond clearly visible individual scars.

The contribution of the method is not limited to recovering scar-like
structures. The same architecture can replace localization with other quantities
carrying physical meaning and learn the elements that best represent a family of
states while optimizing those properties. Neural networks can therefore be used
not only to approximate quantum states, but also as a tool for discovering their
building blocks and relating them to the underlying dynamics.

\begin{acknowledgments}
This work has been partially supported by the Spanish Ministry of Science,
Innovation and Universities, Gobierno de Espa\~na, under Contract
No.~PID2021-122711NB-C21. Support from CONICET is gratefully acknowledged.
\end{acknowledgments}

\appendix

\section{Complex-to-real implementation}
\label{app:complex-real}

For one atom
$a=a_R+\ii a_I$ and state $\psi=\psi_R+\ii\psi_I$,
\begin{align}
 \braket{a}{\psi}
 &=
 (a_R-\ii a_I)^\mathsf{T}(\psi_R+\ii\psi_I)
 \nonumber\\
 &=
 a_R^\mathsf{T}\psi_R+a_I^\mathsf{T}\psi_I
 +\ii\left(a_R^\mathsf{T}\psi_I-a_I^\mathsf{T}\psi_R\right).
\end{align}
Stacking all atoms gives the real block representation
\begin{equation}
 \begin{pmatrix}
  \Re\bm{z}\\
  \Im\bm{z}
 \end{pmatrix}
 =
 \begin{pmatrix}
  A_R&A_I\\
  -A_I&A_R
 \end{pmatrix}
 \begin{pmatrix}
  \psi_R\\
  \psi_I
 \end{pmatrix}.
\label{eq:real-encoder}
\end{equation}
Likewise, synthesis
$\widetilde\psi=A^\mathsf{T}c$ gives
\begin{align}
 \Re\widetilde\psi&=A_R^\mathsf{T}c_R-A_I^\mathsf{T}c_I,\\
 \Im\widetilde\psi&=A_R^\mathsf{T}c_I+A_I^\mathsf{T}c_R.
\end{align}
The conjugation appears in analysis but not in synthesis, exactly as required by
Eqs.~\eqref{eq:analysis} and \eqref{eq:decoder}.

\section{Validation-state identities and displayed pairs}
\label{app:indices}

The three validation examples in Fig.~\ref{fig:reconstruction} have full-spectrum
indices and fidelities
\begin{center}
\begin{tabular}{cc}
\toprule
$n$ & $F_n$\\
\midrule
  & \\
  & \\
  & \\
\midrule
63 & 0.945575\\
208 & 0.943058\\
232 & 0.918748\\
\bottomrule
\end{tabular}
\end{center}
The displayed atom--scar pairs in Fig.~\ref{fig:atom-scar} are
\begin{center}
\begin{tabular}{cccc}
\toprule
$m$ & $r$ & binary string & $O_{mr}$\\
\midrule
 &  &  & \\
 &  &  & \\
 &  &  & \\
\midrule
90 & 69 & 000011 & 0.665872\\
128 & 37 & 00011 & 0.505747\\
147 & 186 & 0001001 & 0.689316\\
\bottomrule
\end{tabular}
\end{center}
Their scar labels additionally carry the quasienergy and symmetry-sector indices
generated by Eqs.~\eqref{eq:scar-quasienergy}--\eqref{eq:symmetries}. These
machine-readable labels should accompany the public data release.

\bibliographystyle{apsrev4-2}
\bibliography{references}

\end{document}